# An Enhanced MPPT Method based on ANN-assisted Sequential Monte Carlo and Quickest Change Detection

Leian Chen and Xiaodong Wang*, *Fellow, IEEE*

### Abstract

The performance of a photovoltaic system is subject to varying environmental conditions, and it becomes more challenging to track the maximum power point (MPP) and maintain the optimal performance when partial shading occurs. In this paper, we propose an enhanced maximum power point tracking (MPPT) method utilizing the state estimation by the sequential Monte Carlo (SMC) filtering which is assisted by the prediction of MPP via an artificial neural network (ANN). A state-space model for the sequential estimation of MPP is proposed in the framework of incremental conductance (I-C) MPPT approach, and the ANN model based on the observed voltage and current or irradiance data predicts the global MPP (GMPP) to refine the estimation by SMC. Moreover, a quick irradiance change detection method is applied, such that the SMC-based MPPT method resorts to the assistance from ANN only when partial shading is detected. Simulation results show that the proposed enhanced MPPT method achieves high efficiency and is robust to rapid irradiance change under different noise levels.

### Index Terms

Photovoltaic (PV) systems, partial shading, maximum power point tracking (MPPT), sequential Monte Carlo, artificial neural network, quickest change detection.

## I. INTRODUCTION

The photovoltaic (PV) technology that harnesses the solar energy to generate electricity has experienced a rapid growth in deployment over the past years. Since the efficiency of power

* Corresponding author.

L. Chen and X. Wang are with the Department of Electrical Engineering, Columbia University, New York, NY, 10027 USA (e-mail: chen.leian@columbia.edu, wangx@ee.columbia.edu).

transfer from the PV cell depends on both the amount of solar irraidance received by PV panels and the electrical characteristics of the load, it is important to maximize power extraction under all conditions. As the amount of solar irradiance varies over time, and so doest the load characteristic that gives the highest power transfer efficiency. The maximum power point tracking (MPPT) technique tracks the load characteristic (called maximum power point (MPP)) that maximizes the power.

*A. Related Works*

MPPT methods can be generally classified into two categories: conventional methods (e.g., the Perturbation and Observation (P&O) method, the Incremental Conductance (I-C) method), and advanced methods (e.g., methods based on fuzzy logic (FL) based, artificial neural network (ANN) based and particle swarm optimization (PSO)).

The conventional MPPT methods are widely used since they are generally simple and cheap. Both P&O [1]–[3] and I-C [4]–[6] methods control the reference signal of a DC-DC converter that matches the PV module voltage with that of the DC bus or works as a battery charge. In the P&O method, the controller adjusts the voltage by a small amount from the array and observes the power change; if the power increases, it adjusts the operating voltage in that direction until the output power no longer increases. The I-C method is based on the fact that the slope of the power-voltage curve characterizing the PV array is zero at the MPP, positive on the left, and negative on the right of the MPP. The controller measures the incremental changes in PV array current and voltage to predict the effect of a voltage change.

However, the conventional methods are not efficient and can even fail under some special conditions, such as an abrupt irradiance change due to shadings. Therefore, more intelligent MPPT techniques have been proposed for better transient and steady-state performance. FL MPPT controllers [7] [8] does not need an accurate mathematical model and can work with imprecise measurement inputs. Variations of the PSO based methods are proposed in [9]–[13], where the controller searches the optima in a population of potential solutions (particles). ANN-based methods [14] [15] have shown good performance under rapidly varying irradiance, especially in terms of efficiency and quick response. [16] presents a reinforcement learning based MPPT method that observes the environment state of the PV array in the training process and autonomously adjusts the perturbation to the online operating voltage.

Note that despite of the higher efficiency, these advanced approaches are much more complex compared to the conventional techniques. Moreover, the performance of the machine learning approaches is heavily dependent on the accuracy of the trained model that is determined by the quality of training data.

*B. Our Contributions*

We aim to find a cost-effective MPPT method by exploiting the advantages of both the conventional and advanced approaches. The contribution of our proposed MPPT method is three fold. 1) To the best of our knowledge, it is the first time that an sequential Monte Carlo (SMC) method is applied under the framework of I-C approach to tackle the nonlinearity when the step-size of voltage adjustment is time-varying. 2) Considering the challenge of partial shading which leads to multiple local optimal operation points, we further adopt the ANN method based on multiple measurement inputs for refinement in the SMC-based I-C method to find the global optimal operation point more efficiently. 3) Moreover, to spare the redundant ANN predictions when the received irradiance is steady, the ANN prediction is triggered only when the proposed GLLR detector declares an irradiance change. The intelligent integration of our SMC-based I-C approach and the refinement by ANN provides an efficient and economical MPPT solution. Extensive simulation results demonstrate that our method is robust to the various shading patterns and process noises.

The remainder of the paper is organized as follows. Section II introduces the mathematical model for PV systems, and the state-of-the-art three-component MPPT framework. In Section III, we present our proposed enhancements to the three-component MPPT approach. In Section IV, the proposed method is applied to a simulated partially shaded PV system and its performance is compared with the existing ANN-based MPPT methods exploiting the observations of the irradiance in [14] or the voltage and current in [15]. Section V concludes the paper.

## II. Background on MPPT for PV Systems

In this section we first give an overview of the structure of an PV system and present the analytical model characterizing the circuits that generate the electricity. We then briefly summarize the three-component framework of MPPT under partial shading adopted in [15] [14] that makes use of ANN to assist the conventional approaches (e.g., I-C and P&O). Finally we highlight our proposed enhancement to each one of the three components of the framework.

## A. Overview of PV Systems and the Analytical Model

PV systems harvest the energy from sunlight to generate electricity via an electronic process in semiconductors. As shown in Fig. 1, a typical PV system for residential or industrial electricity supply usually consists of the PV array which generates electricity directly from sun irradiance, and other follow-up components which are often referred to as "balance of components" (BOC). An $M \times N$ PV array is composed of numerous PV cells which are encapsulated into PV modules. $N$ modules are first serially connected as a string to accumulate the desired output voltage and then $M$ strings are connected in parallel to increase the output current. To utilize the electricity generated by PV arrays, the BOC transforms and stores the energy into the form that can be directly delivered for daily applications. The BOC basically includes the mounting structures to fix and direct PV modules towards the sun, the DC-AC converters (also known as inverters) for applications requiring AC, the MPPT components for adjusting the operating voltage and current, the batteries for energy storage, and a charger regulator for smooth operation of the PV system.

Since the PV array determines the efficiency of transforming the solar energy into electricity, it is of fundamental interest to characterize the relationship between the received irradiance at a given PV cell and the corresponding output power. Various analytical models have been proposed to characterize the circuits in a PV cell.

In this paper, we utilize a typical equivalent single diode circuit model shown in Fig. 2, where $V_{PV}$ and $I_{PV}$ denote the output voltage and current respectively, $R_s$ is the series parasitic resistance, and $R_p$ is the parallel parasitic resistance. Since typically $R_s$ is very small and $R_p$ is very large, they are negligible. The simplified model is given as [17]

$$I_{PV} = I_{ph} - \frac{I_{sc,STC}}{\exp\left(\frac{q}{kT}B\right)} \left[\exp\left(\frac{q}{kT}B\frac{V_{PV}}{V_{oc,STC}}\right) - 1\right], \tag{1}$$

$$I_{ph} = I_{ph,STC} + K_I \left(T - T_{STC}\right) \frac{\lambda}{\lambda_{STC}}, \tag{2}$$

where the definitions of parameters are given in Table I. In this paper, we set $B = 0.2464$ as suggested in [6]. Note that, in (1) and (2), the only variables are the instantaneous irradiance $\lambda$ and temperature $T$, and the corresponding output voltage $V_{PV}$ and current $I_{PV}$; the other parameters are constants determined by the hardware attributes.

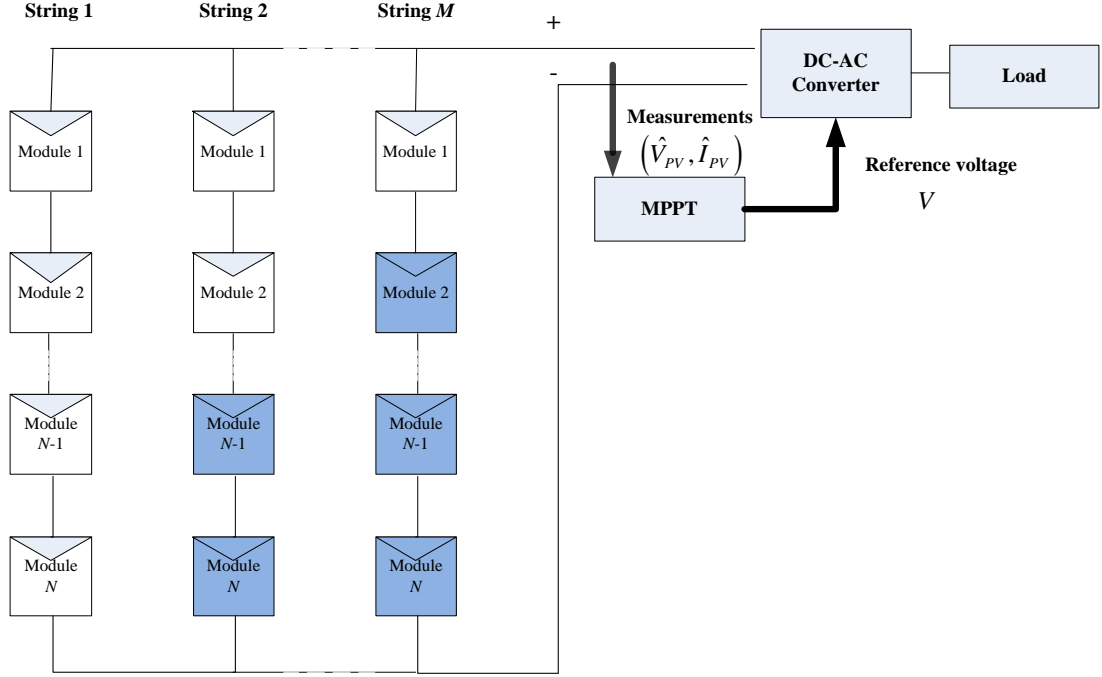

Fig. 1. Illustration of a PV system. The shaded blocks denote the PV module under partial shading.

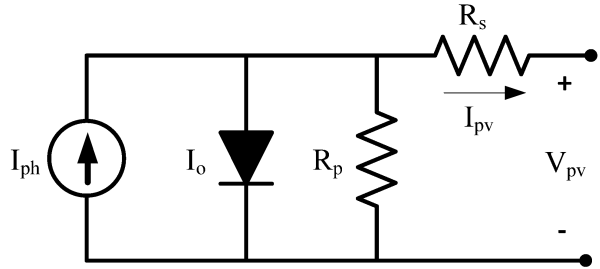

Fig. 2. Equivalent circuit of a PV array.

## B. Three-component MPPT Framework

The task of the MPPT block in a PV system shown in Fig. 1 is to obtain the maximum power output by finding the optimal operating voltage and the corresponding operating current regulated by (1). Specifically, given a certain sampling frequency $f_s$, the MPPT block monitors the output power of a given PV system (including the measured noisy voltage $\widehat{V}_{PV}(t)$ and the current $\widehat{I}_{PV}(t)$) for each time instant $t$, estimates the optimal operating voltage $V(t+1)$ leading to the MPP, and then adjusts the system accordingly to closely follow the MPP under dynamic environmental conditions.

TABLE I
PARAMETERS IN THE SIMPLIFIED PV CIRCUIT MODEL

| Parameter | Definition |
| --- | --- |
| $V_{PV}$, $I_{PV}$ | Output voltage and current of the PV array |
| $I_{ph}$, $I_{ph,STC}$ | Photo-generated current under operation and standard test conditions (STC), i.e., at 1kW/m$^2$ and 25°C |
| $I_{sc,STC}$ | Short circuit current measured at STC |
| $T$ | Temperature under operation |
| $V_{oc}$, $V_{oc,STC}$ | Open circuit voltage under operation and STC |
| $\lambda$, $\lambda_{STC}$ | Irradiance under operation and STC in kW/m$^2$ |
| $K_I$ | Temperature coefficient of short-circuit current |
| $q$ | Electron charge |
| $k$ | Boltzmanns constant |

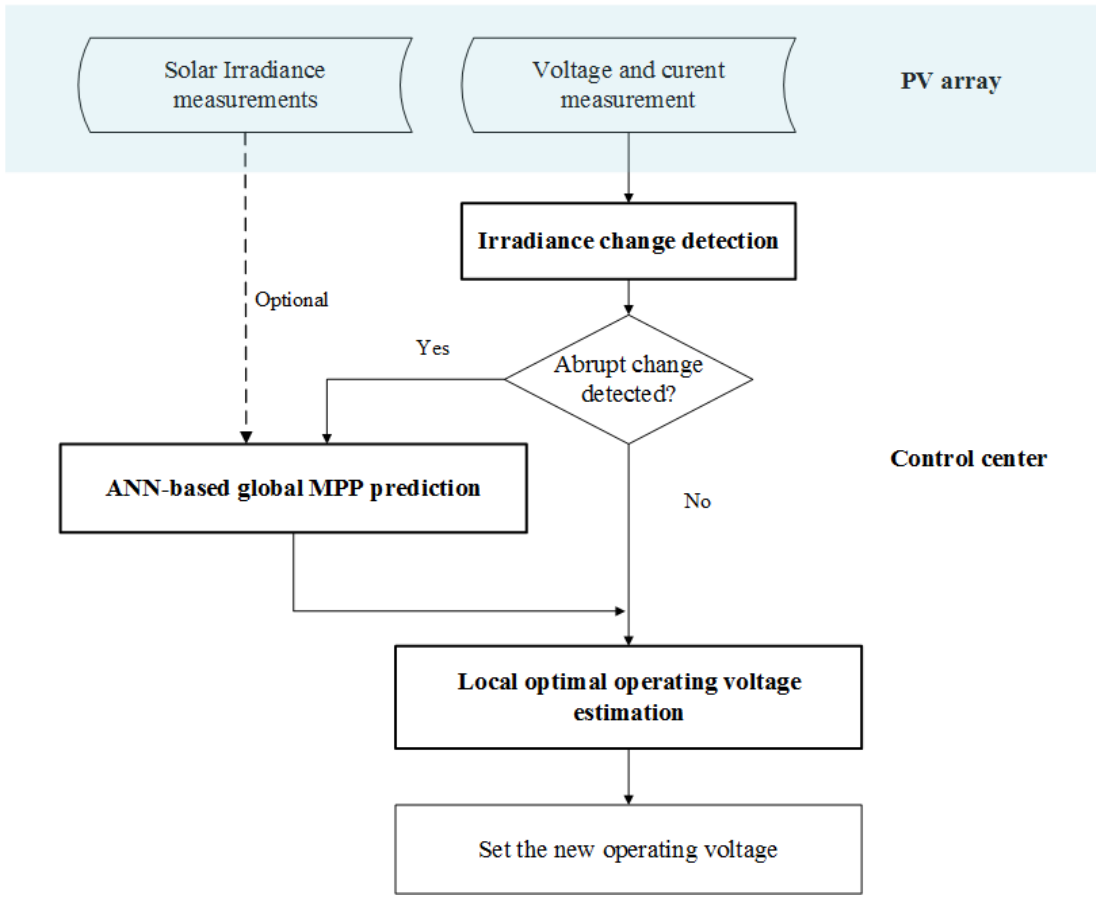

Fig. 3. Flowchart of the three-component MPPT method.

Conventional methods (P&O, I-C, etc.) only relies on the analytical model in (1)-(2) and the observation data to determine the next adjustment, and they tend to fail when the PV system is subject to rapid environmental change. Recent works have proposed a three-component MPPT framework illustrated in Fig.3 to overcome the drawbacks of conventional approaches by making use of machine learning tools [14] [15] [16]. Generally, the three components tackle the problems of the detection of partial shading, (local) optimal operating voltage estimation based on conventional methods and the prediction of the global MPP (GMPP) using machine learning methods, respectively. In particular, in [14] [15], given an online measurement at time $t$, the controller decides whether the ANN prediction needs to be triggered due to an abrupt irradiance change. If so, the prediction from ANN will refine the estimation of the operating voltage at $t+1$ via conventional methods; otherwise, only the conventional I-C or P&O methods is implemented under the assumption of steady environmental conditions. Finally, the controller directs the system to operate at the estimated voltage and observes the output power at $t+1$.

*C. Proposed Enhancements*

In this paper, we also adopt the three-component framework but we propose new method for each component to enhance the overall MPPT performance.

1) For irradiance change detection, the simple threshold-based detection rule in [14] only relies on the difference between two consecutive power measurements and tends to have large detection delay when a long false alarm period is desired. Since the power readings are not constant due to measurement noise, we need to distinguish the abnormal fluctuation due to partial shading or other faulty conditions from the noisy measurements, and quickly detect any small faulty signal given a certain false alarm period. In this work, we propose to employ the sequential change detection technique for this purpose.

2) For the optimal operating voltage estimation, [14] [15] simply use an I-C method with the kernel of voltage transition model given by

$$V(t+1) = V(t) + m_0 \frac{dP(t)}{dV(t)}, \tag{3}$$

where $m_0$ is a step-size constant, and $\frac{dP(t)}{dV(t)}$ is the slope of the power-vs-voltage (P-V) curve illustrated in Fig. 4. On one hand, the controller does not adjust the operating voltage when $\frac{dP(t)}{dV(t)} = 0$ indicating the peak in the P-V curve. However, the efficiency can be further improved by using a variable step-size. Intuitively, given the prediction of the GMPP, we can use a relatively large step-size when the current operating voltage is far from the GMPP and a small step-size otherwise. On the other hand, note that the output current $I_{PV}$ and the voltage $V_{PV}$ in (1) are

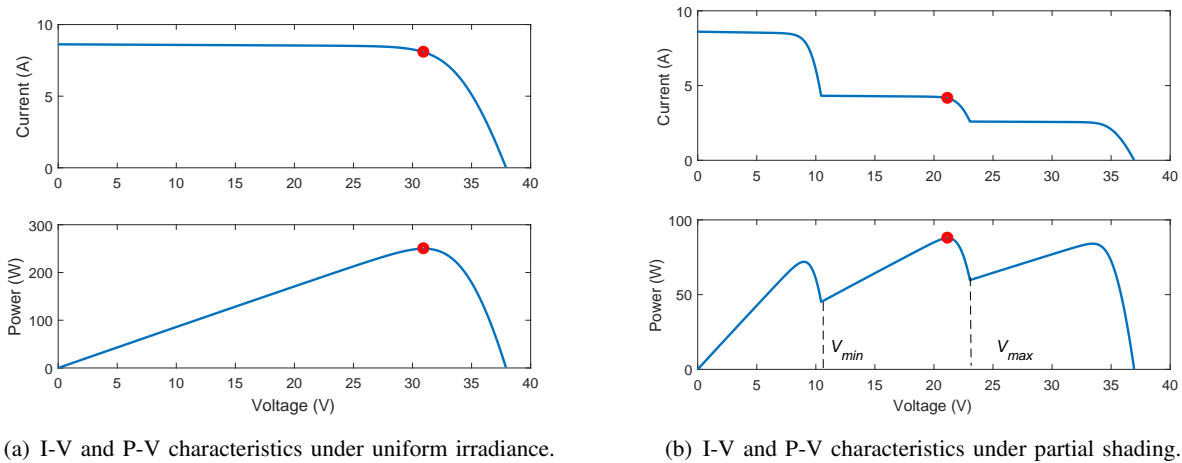

(a) I-V and P-V characteristics under uniform irradiance.

(b) I-V and P-V characteristics under partial shading.

Fig. 4. I-V and P-V characteristics of a 3-panel PV system under different irradiance conditions. The red dot denotes the global MPP point

assumed to be noise-free while only noisy measurements can be obtained in real PV systems and the predicted $V(t+1)$ in (3) is not the true (noise-free) optimal operating voltage. Since we can never tell if the MPP is achieved given a single noisy measurement, we need to sequentially estimate the reference voltage $V(t)$ (a noise-free state variable) which leads to MPP at each sampling instant $t$. The controller iteratively tunes the operating voltage given the current estimate of $V(t)$, and updates the estimate with new measurements at $t+1$. When the voltage transition model becomes nonlinear as we adopt a variable step size, the voltage estimation based on the basic I-C method is no longer effective, which necessitates the adoption of nonlinear estimation approach.

3) To tackle the MPPT under partial shading when multiple peaks occur in the P-V curve as in Fig. 4(b), the ANN is used in [14] [15] to predict the GMPP and forces the I-C estimation within in the region of GMPP, $(V_{min}, V_{max})$. In this paper, we adopt an ANN model with the input data from a short sequence of voltage and current measurements or several irradiance measurements from different PV panels. In comparison with [14] and [15] where only the instantaneous total voltage and current or irradiance measurements are used, the prediction of our ANN model is more robust especially when the measurements are noisy.

## III. PROPOSED ENHANCED MPPT

Based on the above analytical system model, an improved MPPT method under partial shading is developed. We first give an overview of the proposed method and then specify the detailed steps.

Following the three-component framework in Fig. 3, we propose to replace each component by some more advanced module. In particular, for the irradiance change detection component, we propose to employ the sequential GLLR change detection; for the nonlinear local MPP estimation, we adopt the sequential Monte Carlo method; for the GMPP prediction, a trained ANN model based on multiple local measurements is triggered when an irradiance change occurs. The overall MPPT block works as follows. First, the measurements (e.g., power, voltage, current, irradiance) are acquired by the MPPT controller. Whenever the proposed GLLR change detector does not declare an abrupt power variation, we assume that the P-V curve maintains the same shape as in the previous sampling cycle, and the MPP is tracked by the SMC-based method. If the change detector declares an abrupt power fluctuation due to partial shading, the ANN-assisted MPPT algorithm is triggered. The trained ANN model utilizes a small group of several local measurements as the input and predicts the GMPP under the current shading condition. Then, refined by the prediction of ANN, the SMC-based I-C algorithm updates the voltage estimation. Finally, the controller steers the operating point to the estimated optimal voltage $\widehat{V}(t)$ and waits for the next MPPT cycle.

In the following subsections, we first present the GLLR-based quickest change detection method, then describe the voltage state estimation by SMC given the input from the ANN, and finally specify the GMPP prediction method by ANN.

### A. GLLR Change Detector

Conventionally, the MPPT is continuously run whenever the PV system is under operation, which is not economical since normally the output power does not exhibit dramatic changes with steady irradiance input during which the MPPT is redundant. Existing works have proposed some simple rules to trigger the ANN prediction in MPPT. For example, in [14], the ANN is implemented to assist the conventional P&O and I-C methods when the difference between adjacent power measurements $|P(t) - P(t-1)|$ exceeds a predetermined threshold. Such a method is quite ad hoc and is not effective in fast detection of partial shading. Here we propose a

principled approach to detecting the abnormal fluctuation in the framework of sequential quickest change detection [18].

In what follows, we first formulate the problem of sequential irradiance change detection based on the vector autoregressive (AR) model. Then a quickest fault detection method based on the GLLR test is presented. We assume that the statistical properties of the measurement signal (voltage, current or power) under normal conditions, i.e., before the irradiance change, are known (or available from historical data) while properties of the signal after the change are totally unknown.

*1) Power Measurement Models:* Our goal is to detect an unknown power shift between different irradiance conditions. We assume that the output power of a PV system before an abrupt irradiance change is known. Specifically, when the system is under uniform irradiance with a certain set of environmental inputs (irradiance, temperature, etc.), the DC output power should be a constant known by design; when the system is under a relatively steady shading condition, the corresponding nominal output power can be approximated by the average meter readings. Then we can subtract the constant from the measured voltage signal to obtain the post-processing signal $\widetilde{P}(t)$ that comprises the measurement noise only or possibly a "faulty" signal due to irradiance change. In particular, before the power shift occurs at $t = t_0$, the post-processing signal consists of the measurement noise $\nu(t)$ only, which is modeled as a white Gaussian process, i.e.,

$$\widetilde{P}(t) = \nu(t) \sim \mathcal{N}(0, \sigma_\nu^2), \qquad t < t_0, \tag{4}$$

where $\sigma_\nu^2$ is the variance.

After $t_0$, the DC power output is corrupted by both the faulty signal and noise, i.e.,

$$\widetilde{P}(t) = s(t) + \nu(t), \quad t \geq t_0, \tag{5}$$

where $s(t)$ is the disturbance signal caused by the irradiance change. Different from the measurement noise $\nu(t)$ that is uncorrelated in time, the faulty signal $s(t)$ is correlated in time, which is the basis for the GLLR detection scheme in this paper. In particular, we use an autoregressive (AR) model to characterize the statistical property of the faulty signal $s(t)$. The AR model has been employed to characterize the disturbance [19] [20] and inter-area oscillations [21] in power grid systems, the gear tooth fault signals in mechanical systems [22], speech signals [23], etc.

Specifically, the faulty signal in the PV system is modeled as

$$s(t) = \tilde{\mu} + \sum_{j=1}^{p} a(j)\left[s(t-j) - \tilde{\mu}\right] + \omega(t) \tag{6}$$

where $p$ denotes the order of the AR model, $\tilde{\mu}$ is the mean, $\omega(t)$ is the innovation noise and $a(j)$, $j = 1, ...p$, are the AR coefficients. Substituting (6) into (5), we have

$$\widetilde{P}(t) = \mu + \sum_{j=1}^{p} a(j)\widetilde{P}(t-j) + \varepsilon(t), \quad t \geq t_0 \tag{7}$$

where $\mu \triangleq \mu[1 - \sum_{j=1}^{p} a(j)]$ and $\varepsilon(t) = [\nu(t) - \sum_{j=1}^{p} a(j)\nu(t-j) + \omega(t)] \sim \mathcal{N}(0, \sigma_\varepsilon^2)$ with $\sigma_\varepsilon^2 = [1 + \sum_{j=1}^{p} a^2(j)]\sigma_\nu^2 + \sigma_\omega^2$ reflecting the impacts of both the disturbance signal and the measurement noise.

*2) GLLR Change Detector:* The occurrence of an power change of interest is declared at time $T$ via the following sequential change detection procedure, called the generalized local likelihood ratio (GLLR) test [18],

$$g_t = (g_{t-1} + l_t)^+, \tag{8}$$

$$T = \inf\{t : g_t \geq h\}, \tag{9}$$

where $(x)^+ \triangleq \max\{x, 0\}$, $g_0 = 0$, $h$ is a threshold determined by the desired false alarm period $\gamma$, and $l_t$ is the generalized log likelihood ratio (GLLR) given all previous measurements by $t$. As in [24], we aim to detect a small change corresponding to the most challenging scenario, and thus assume that $\boldsymbol{\theta}_0 \approx \boldsymbol{\theta}_1$ where $\boldsymbol{\theta}_0$ and $\boldsymbol{\theta}_1$ denotes the model parameters before and after change. To that end, $l_t$ is approximated by the local second-order expansion of the GLLR by assuming $\boldsymbol{\theta}_1 \to \boldsymbol{\theta}_0$, given by

$$l_t \approx b \|\tilde{\boldsymbol{z}}_t\| - \frac{1}{2}b^2, \tag{10}$$

where $\|\cdot\|$ denotes the vector $l_2$-norm, $b > 0$ is a predetermined parameter reflecting the deviation of the output power, and

$$\tilde{\boldsymbol{z}}_t = \begin{bmatrix} \frac{1}{\sigma_\nu^2}\widetilde{P}(t)\widetilde{\boldsymbol{P}}_{t-p}^{t-1} \\ \frac{1}{\sqrt{2}}\left(\frac{\widetilde{P}(t)^2}{\sigma_\nu^2} - 1\right) \\ \frac{\widetilde{P}(t)}{\sigma_\nu} \end{bmatrix}, \tag{11}$$

where $\widetilde{\boldsymbol{P}}_i^j \triangleq \left[\widetilde{P}(i), \widetilde{P}(i+1), ..., \widetilde{P}(j)\right]$.

Note that whenever $g_t$ exceeds the threshold $h$ at time $T$, the detector declares an abnormal power change and the ANN-assisted MPPT algorithm is implemented. Otherwise, we implement the SMC-based MPPT without the GMPP estimation by ANN, assuming the P-V curve does not change when no irradiance change occurs. The GLLR detector is restarted immediately after the previous alarm and the post-processing signals during this detection cycle are obtained by subtracting the new power measurements at time $T+1$. Generally, smaller values of $b$ and $h$ leads to a more "sensitive" detector but a false alarm could occur when the irradiance condition actually does not change and the ANN is triggered unnecessarily. The value of $b$ is generally set to be proportional to the average drifts of empirical voltage readings (i.e., noise level ) when the PV system is under normal operation. Since a large false alarm period is usually preferred, the threshold $h$ is tuned accordingly.

## B. Voltage State Estimation via SMC

We first present a nonlinear state-space model characterizing the voltage estimation process, and then specify the SMC algorithm based on the defined model.

*1) State-space Model:* To adopt the framework of SMC, we define the voltage state transition model and the measurement model as follows. A discrete-time state-space model is considered for a PV system, i.e. the sampling frequency $f_s$ is a fixed constant. Based on the I-C structure in (3), we define the reference voltage $V(t)$ as the state variable updated over time and the noisy voltage readings $\widehat{V}_{PV}(t)$ as the measurement variable. Note that $V(t)$ is the hypothesized quantity reflecting the estimated GMPP and the operating voltage of the PV system is adjusted to get close to $V(t)$.

- **State Transition Model**

The P-V curve illustrates that the power increases with a gradual positive slope until it reaches a global/local optimal point, and then decreases sharply after that, which inspires the so called incremental conductance (I-C) MPPT method. The I-C method is based on the observation that at the maximum power point, the slope $dP/dV_{PV} = 0$ and $P = I_{PV}V_{PV}$. Accordingly, a general

form of the voltage estimation model in the I-C approach is given as

$$V(t+1) = f(V(t)) + u(t) + w(t)$$
$$= \left[V(t) + m(t)\frac{dP(t)}{dV_{PV}(t)}\right] + u(t) + w(t), \quad (12)$$

where $P(t)$ and $V_{PV}(t)$ are the true noise-free power and voltage observations; $m(t)$ determines the step size of the voltage adjustment, $u(t)$ is an external input for refinement (e.g., ANN prediction result), and $w(t) \sim \mathcal{N}(0, \sigma_w^2)$. Given (1), the true instantaneous slope $\frac{dP(t)}{dV_{PV}(t)}$ is expressed by

$$\frac{dP(t)}{dV_{PV}(t)} = \frac{dV_{PV}I_{PV}(t)}{dV_{PV}(t)} = I_{PV}(t) + V_{PV}\frac{dI_{PV}(t)}{dV_{PV}(t)}$$
$$= I_{PV}(t) - \frac{V_{PV}(t)I_{sc,STC}}{\exp(\frac{qB}{KT})}\left[\frac{qB}{KTV_{oc,STC}}\exp\left(\frac{qBV_{PV}(t)}{KTV_{oc,STC}}\right)\right]. \quad (13)$$

Note that, in practice, the value of $\frac{dP(t)}{dV_{PV}(t)}$ can be either directly reported by meters or computed given the observed voltage and current (and thus are noisy). For the latter case, we can use the approximate value $\frac{d\widehat{P}(t)}{d\widehat{V}_{PV}(t)}$.

In particular, we propose to employ an adaptive step size $m(t)$ given as

$$m(t) \triangleq m_0[V(t) - V_{EGMPP}(t)]^2, \quad (14)$$

in contrast with the conventional methods where the step size is a constant $m_0$. The proposed $m(t)$ ensures that the step size is relatively small when the current reference voltage is close to the estimated GMPP $V_{EGMPP}$, and thus increases the "resolution" of the I-C method. The constant parameter $m_0$ can be chosen by fitting a sequence of voltage measurements during the MPPT under uniform irradiance to the model in (12) based on the chosen criterion, e.g., the least squares method.

The refinement function $u(t)$ compensates the gap between the current noisy voltage measurement $\widehat{V}_{PV}(t)$ and the latest estimated GMPP $V_{EGMPP}(t)$ by ANN, given as

$$u(t) \triangleq \left[V_{EGMPP}(t) - \widehat{V}_{PV}(t)\right] \mathbb{1}_{g_t > h}. \quad (15)$$

(15) implies the refinement term is nonzero only when an irradiance change is declared. Note that the refinement term $u(t)$ effectively drives the system to operate at the global optimum in case that the system gets stuck at a local optimum.

**Algorithm 1** SMC-based State Estimator
─────────────────────────────────────────────
1: Initialization: $t = 0$, draw $N$ samples, $\{V^{(j)}(0)\}_{j=1}^{N}$, according to the prior distribution $\mathcal{N}(V_0, \sigma_0^2)$. Set the weight $w^{(j)}(0) = 1/N$ for all $j$.
2: **for** $t = 1, 2, ...$ **do**
3:     Generate the current samples $V^{(j)}(k)$ according to (19).
4:     Update the weight $w^{(j)}(k)$ according to (20)
5:     Compute the state estimates, $\widehat{V}(t)$, according to (22).
6:     Perform resampling if $\widehat{N}_{\text{eff}}$ is below a given threshold.
7: **end for**
─────────────────────────────────────────────

- **Measurement Model**

Since the measured voltage includes the true reference voltage and the noise, the measurement equation is given by

$$\widehat{V}_{PV}(t) = V(t) + v(t), \tag{16}$$

where $v(t) \sim \mathcal{N}(0, \sigma_v^2)$ is a Gaussian measurement noise.

Given the input values, $\widehat{V}_{PV}(t)$, $\frac{dP(t)}{dV_{PV}(t)}$ and $u(t)$, an SMC filtering can be implemented to estimate $V(t)$ based on the nonlinear state transition model (12) and the measurement model (16).

*2) Online Estimation by SMC:* Our proposed MPPT approach is based on the voltage estimation by the SMC method. In SMC [25], a set of weighted samples are used to approximate an underlying distribution that is to be estimated. And the samples and their associated weights are sequentially updated based on the new measurements. The proposed SMC-based state estimator is summarized in Algorithm 1. The algorithmic details are presented as follows.

- **Initialization**

For $t = 0$, draw $N$ initial samples, $\{V^{(j)}(0)\}_{j=1}^{N}$, from the prior probability density function characterized by $V_0$ and $\sigma^2{}_0$, i.e.,

$$V^{(j)}(0) \sim \mathcal{N}(V_0, \sigma_0^2), \quad j = 1, 2, ..., N. \tag{17}$$

Set initial weights $w^{(j)}(0) = 1/N$ for all $j$. Note that the prior distribution $\mathcal{N}(V_0, \sigma_0^2)$ for drawing the initial samples can be estimated during the offline training process. Specifically, given a set of voltage measurements under normal operation with uniform irradiance, $V_0$ and $\sigma_0^2$ are approximated by the sample mean and variance respectively. The number of samples $N$ should be sufficiently large to characterize the distribution of $V(t)$. Numerical studies reveals that $300 < N < 800$ can provide decent estimation accuracy given the computational budget.

- **Online State Estimation**

During the online phase, the MPPT controller sequentially updates the state estimate and adjusts the operating voltage accordingly. An SMC update step consists of *sample generation*, *weight update*, and *resampling*, as highlighted in Algorithm 1.

*Sample generation*: The basic idea of SMC is to perform the sequential importance sampling (SIS). At each time, $N$ samples $\{V^{(j)}(t)\}_{j=1}^{N}$ are drawn from some trial distribution $\pi\left(V^{(j)}(t)|\boldsymbol{V}^{(j)}(t-1), \widehat{\boldsymbol{V}}_{PV}(t)\right)$ with $\boldsymbol{V}^{(j)}(t-1) \triangleq \{V^{(j)}(1), V^{(j)}(2), ..., V^{(j)}(t-1)\}$ and $\widehat{\boldsymbol{V}}_{PV}(t) \triangleq \{\widehat{V}_{PV}(1), \widehat{V}_{PV}(2), ..., \widehat{V}_{PV}(t)\}$. Here we choose the state transition density as the trial distribution, i.e.,

$$\pi\big(V^{(j)}(t)|\boldsymbol{V}^{(j)}(t-1), \widehat{\boldsymbol{V}}_{PV}(t)\big) \triangleq p\big(V^{(j)}(t)|V^{(j)}(t-1)\big). \tag{18}$$

Hence according to (13),

$$V^{(j)}(k) \sim \mathcal{N}\left(f\left(V^{(j)}(t-1)\right) + u(t), \sigma_w^2\right). \tag{19}$$

*Weight update*: The corresponding weight $w^{(j)}(t)$ for sample $V^{(j)}(t)$ is calculated by

$$\begin{aligned}
w^{(j)}(t) &\propto \overline{w}^{(j)}(t-1) \frac{p(\widehat{V}_{PV}(t)|V^{(j)}(t))p(V^{(j)}(t)|V^{(j)}(t-1))}{\pi\left(V^{(j)}(t)|\boldsymbol{V}^{(j)}(t-1), \widehat{\boldsymbol{V}}_{PV}(t)\right)} \\
&\propto \overline{w}^{(j)}(t-1) p\left(\widehat{V}_{PV}(t)|V^{(j)}(t)\right) \\
&\propto \overline{w}^{(j)}(t-1) \cdot \exp\left[-\frac{1}{2\sigma_v^2}\left(\widehat{V}_{PV}(t) - V^{(j)}(t)\right)^2\right],
\end{aligned} \tag{20}$$

where the normalized weight $\overline{w}^{(j)}(t)$ is given as

$$\overline{w}^{(j)}(t) = \frac{w^{(j)}(t)}{\sum_{j=1}^{N} w^{(j)}(t)}. \tag{21}$$

Given the current weighted samples $\{V^{(j)}(t), \overline{w}^{(j)}(t)\}_{j=1}^{N}$, we can estimate the state variable as

$$\widehat{V}(t) = \sum_{j=1}^{N} \overline{w}^{(j)}(t) V^{(j)}(t). \tag{22}$$

Then the controller sets the operating voltage to $\widehat{V}(t)$.

*Resampling*: The resampling step aims to avoid the problem of degeneracy of the SMC algorithm, that is, the situation that all but one of the importance weights are close to zero

[26] [27]. The basic solution is to retain the samples with high weights and discard the samples with low weights.

The resampling is implemented only when the effective number of samples $N_\text{eff}$ is below a predetermined threshold $N_\text{thr}$. An estimate of $N_\text{eff}$ is given by

$$\widehat{N}_\text{eff} = \frac{1}{\sum_{i=1}^{N}\left(\overline{w}^{(j)}(t)\right)^2}, \tag{23}$$

which reflects the variation of the weights [26]. If $\widehat{N}_\text{eff}$ is less than a given threshold, $N_\text{thr}$, we perform resampling to obtain $N$ new samples, $\left\{\widetilde{V}^{(j)}(t)\right\}_{j=1}^{N}$, that is, to draw $N$ samples from the current sample set with probabilities proportional to the corresponding weights.. The corresponding weights for the new samples are set as $\widetilde{w}^{(j)}(t) = 1/N$.

### C. ANN-assisted Maximum Power Point Prediction

As illustrated in Fig. 4, when partial shading occurs, the P-V curve has multiple peaks, and the basic I-C approach without the refinement term $u(t)$ can mislead the system to operate at the local optimal points when the initial estimation falls out of the global optimal region $(V_{min}, V_{max})$. To tackle the disadvantage of the I-C approach, a trained artificial neural network is applied to estimate the global optimum when an irradiance change is detected and thus forces the I-C estimation to fall into the region of GMPP. In particular, we propose to adopt the architecture and inputs specified as follows.

*1) ANN Architecture for MPPT:* In this paper, we apply the most popular ANN structure, a multilayered feed-forward neural network (FNN) [28], which consists of an input layer, one or more hidden layers, and an output layer. A neuron is a processing unit that first linearly weights the inputs, then feeds the sum to a nonlinear activation function $g(.)$, and propagates the results to the following neurons [29]. Explicitly, denoting the $K$ incoming signals at a given node $j$ as $\{x_{1,j}, x_{2,j}, ..., x_{K,j}\}$ with the corresponding weights $\{\omega_{1,j}, \omega_{2,j}, ..., \omega_{K,j}\}$, the output $y_j$ of node $j$ is given as

$$y_j = g\left(\sum_{i=1}^{K} \omega_{i,j} x_{i,j} + \alpha\right), \tag{24}$$

where $\alpha$ is a bias value, and the weights associated with the input values are adjusted by the learning rule in the training process. Various activation functions have been proposed, and we utilize the sigmoid functions [29] given as

$$g(z) = \frac{1}{1+\exp(-z)}, \quad (25)$$

where $z \triangleq \sum_{i=1}^{K} \omega_{i,j} x_i + \alpha$.

*2) Input Variables of ANN:* In particular, we consider a multilayer FNN where the neurons of the input layer simply act as buffers for distributing the input signal and the number of hidden layers is determined in the training phase. For MPPT, the input signals can be PV array parameters (e.g, PV voltages and currents), environmental data (e.g., irradiance and temperature), or any combination of these. The output is usually one or several reference signal(s), such as the estimated GMPP which is used to drive the electronic converter to operate at or close to the MPP. To establish an ANN model for online MPPT, we need to train the model utilizing the input and output data from empirical measurements or model-based simulation results.

Here we present two ANN methods for predicting the global optimum using different input data i.e., voltage and current, and irradiance measurements.

- **ANN based on Voltage and Current Measurements**

When the ANN-assisted MPPT is triggered at $t$, the PV panels are forced to operate $M$ different voltages by tuning the resistance observed by the PV system at the next $M$ sampling time instants. Essentially, the choice of $M$ online inputs can be randomly generated as along as these values fall into the range of possible operating voltages since the training data for the ANN model should cover the whole scope. To utilize the current online observation, we propose to independently generate $M$ inputs in the interval $\left[\widehat{V}_{PV}(t) - 10, \widehat{V}_{PV}(t) + 10\right]$. Then the ANN acquires the output voltage measurements along with the corresponding currents as the input of ANN. In a word, the $M$ online samples $\{(V_{PV,i}, I_{PV,i})\}_{i=1}^{M}$ are fed to ANN, and the estimated voltage $V_{EGMPP}$ corresponding to the global MPP given the current operating condition is the output of the ANN, which is used to refine the estimation by SMC in (15).

- **ANN based on Irradiance Measurements**

Owing to the hardware that supports online irradiance measurements, we can directly model the relationship between the received solar energy and the output power. The measured irradiance data at different locations among PV panels serves as the inputs of ANN, i.e., the number of input neurons $M$ equals the number of available local irradiance measurements in a given system. Again, the output of the ANN is the estimated optimal voltage $V_{EGMPP}$.

*3) Training and Selection of ANN models:* After setting the number of inputs $M$, the number of hidden layers, and the number of neuron in the each hidden layer, we can train the given ANN model during the offline phase. First, we need to obtain a set of training patterns (including the measured voltage and current or irradiance and the GMPP obtained via the P-V curve) that covers the possible operating conditions. For instance, in a $x \times y$ PV array, we manually adjust the operating voltage and current or the input irradiance of each panel by choosing from $z$ possible values, and record the corresponding output GMPPs. Then we have totally $z^{x \times y}$ pieces of training data. Then $\{\omega_{i,j}\}$ and $\alpha$ can be obtained based on the training samples, e.g., by using the back-propagation (BP) algorithm with the Levenberg-Marquardt optimization method [14]. In general, the training algorithm is used to find the weights that minimize some overall error measure such as the mean squared error (MSE). Since even with the same training data, the performance varies when we choose difference ANN models, we need to find an "optimal" one before the online implementation.

To evaluate the performances of different ANN models, we first define a key index quantifying the prediction accuracy.

- **Prediction Quality Index (PQI)**

Generally, the accuracy of the ANN model can be improved when the number of hidden layers and the number of neurons increase, at the cost of computational complexity. To assess the prediction accuracy of a given ANN model, we define the average prediction quality index (PQI) as [14]

$$PQI = \frac{1}{G} \sum_{i=1}^{G} \left( \frac{V_{EGMPP,i}}{V_{GMPP,i}} \right) \times 100\%, \tag{26}$$

where $G$ is the number of ANN prediction tests, $V_{EGMPP,i}$ is the estimated GMPP in the $i^{th}$ test, and $V_{GMPP,i}$ is the true GMPP under the corresponding irradiance condition.

- **Performance Evaluation based on PQI**

Here we present the training results of an irradiance-based ANN as an illustrative example. In our simulated PV system with $3 \times 3$ modules specified in Table II, $6^{3 \times 3}$ training samples are available when the irradiance of each cell is chosen from 6 possible values.

In order to evaluate the accuracy performances of the ANN prediction, different ANN structures have been investigated. In particular, Table III reports the variation ranges related to the number of input PV couples, the number of hidden layers and neurons in the first hidden layer (HL1). The number of neurons in the hidden layers following the first one is set as half of neuron number in the previous layer. During the training process, each ANN model was trained by using $6^{3\times 3}$ data points. In particular, to apply the BP, the input (i.e., values of ($V_{PV}, I_{PV}$) or the irradiance) and the output (i.e., estimated voltage leading to GMPP) obtained from the P-V characteristic curve have been provided to each ANN model 500 times.

TABLE II
SPECIFICATION OF PV MODULES

| Parameter | Value |
|---|---|
| $V_{oc,STC}$ | 21.06 V |
| $I_{sc,STC}$ | 3.80 A |
| Current at $P_{max}$ ($I_{MPP}$) | 3.50 A |
| Voltage at $P_{max}$ ($V_{MPP}$) | 17.10 V |
| Maximum power ($P_{MPP}$) | 59.90W |
| $V_{oc}$ coef. of temperature ($K_V$) | 0.084 V/C |
| $I_{sc}$ coef. of temperature ($K_I$) | 3.3 $\times 10^{-4}$ A/C |

TABLE III
PARAMETER RANGES OF ANN MODELS

| Parameter | Min. | Max. |
|---|---|---|
| Number of inputs (($V_{PV}, I_{PV}$) or irradiance) | 3 | 10 |
| Number of hidden layers | 1 | 4 |
| Size of the first hidden layer | 4 | 20 |

With the trained ANN models, the PQIs were evaluated by simulating 1000 randomly-generated irradiance inputs for each ANN model. For each random experiment, the PQI was calculated based on the reported $V_{EGMPP}$. Fig. 5(a)-(c) are the plots of the Pareto frontier as a scatter plot matrix which represents a 3-D Pareto frontier by showing the value of a pair of objective functions in each figure for each solution. Particularly, Fig. 5(a) and 5(b) demonstrate the tradeoff between the number of input pairs and synapsesa gainst the values of PQI, respectively. That is, larger numbers of input pairs and synapses lead to higher prediction accuracy.

## IV. SIMULATION RESULTS

In this section, we present the performance of the proposed method in both small and large simulated PV systems. For each system with different shading patterns (SP), we first show the performance improvement in each of the three components by comparing with the counterparts in [15] [14], and then evaluate the overall enhancement by presenting the power tracking dynamics of our proposed method in comparison with the methods in [15] and [14]. To quantify the overall

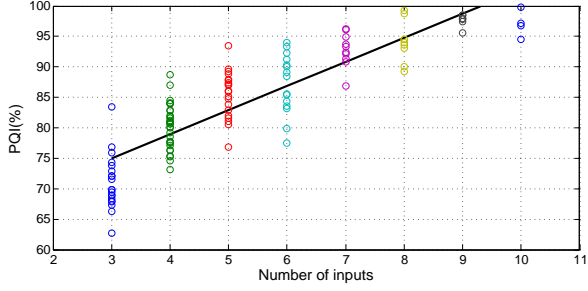

(a) Number of inputs versus PQI.

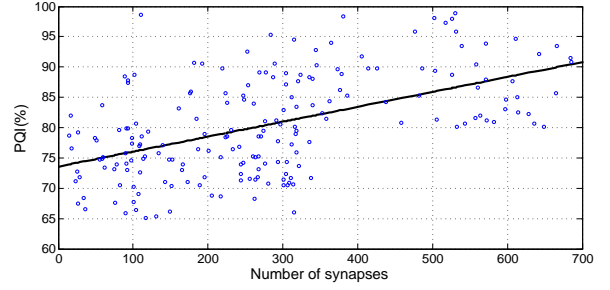

(b) Number of synapses versus PQI.

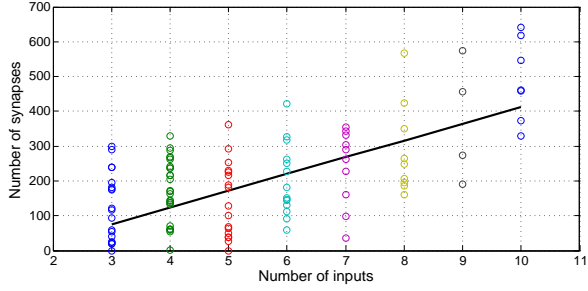

(c) Number of inputs versus number of synapses.

Fig. 5. Scatter plot matrix of the Pareto frontier.

efficiency of the proposed MPPT algorithm, we run the proposed MPPT method 500 times with the identical setup, and present the ratio $\frac{\overline{V}_{PV}(t)}{V_{GMPP}}$ for $T_{shading} \leq t \leq T_{GMPP}$, where $\overline{V}_{PV}(t)$ is the average observed output voltage at $t$, $T_{shading}$ is the time instant when the shading occurs, and $T_{GMPP}$ is the instant when the system arrives at the new GMPP.

### A. System Setup

In our experiments, we simulate different shading conditions in a PV system using the toolbox Simscape Power Systems in Matlab. As illustrated in Fig. 6, the number of PV cells and the input irradiances in each PV array can be tuned to simulate different shading conditions. The input irradiance, and the output power, voltage and current fed into the MPPT algorithm are obtained from the simulated system. The detailed setups to implement the proposed method and other methods for comparison are described below.

*1) Proposed Method:* The MPPT controller adjusts the operating voltage every 0.05s, i.e., the sampling frequency $f_s = 20$ Hz. The variance of the process noise $\sigma_v^2$ is set to $10^{-5}$ (small PV system) or $10^{-3}$ (large PV system). The measurement noise variance $\sigma_w^2 = 10^{-3}$ for both systems. For the GLLR change detector, the threshold $h$ is tuned based on the desired false alarm period $\lambda$. The order of the AR model is set as $p = 5$. In particular, to evaluate the performance of the AR model in characterizing faulty signals, we employed the cross-validation method [30] [24] and compared the model predicted output with the actual output. The numerical results show that the normalized root-mean-squared error (NRMSE) of the AR prediction varies from 12% to 2% for when the order $p$ increases from 1 to 5, which demonstrates the high effectiveness of the AR model in characterizing the faulty signals in PV systems.

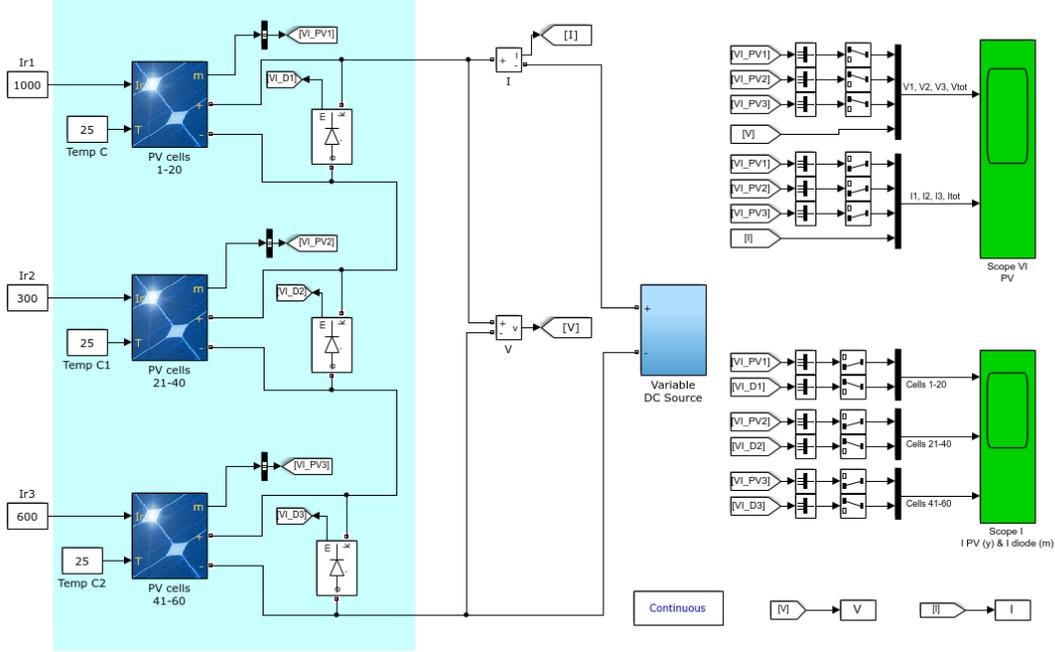

Fig. 6. A simulated PV system under partial shading (four PV panels).

For the SMC algorithm, the step-size coefficient $m_0 = 10^{-2}$, the number of samples before and after resampling are set as $N = 500$. The initial samples $\{V^{(j)}(0)\}_{j=1}^{N}$ are drawn from the prior probability density function $\mathcal{N}(V_0, \sigma_0^2)$ where $V^0$ is the nominal voltage output under normal operation, and $\sigma_0^2$ is approximated by the sample voltage variance obtained during the offline phase.

For the ANN models, the number of inputs is 8 for both the voltage-and-current-based and the irradiance-based model, and the number of hidden layer is 2 each with size 20 and 10.

*2) Methods for Comparison:* Here we briefly describe the methods in [15] and [14] where different ANN models are integrated with the basic I-C approach. The step size of tuning the reference voltage $V(t)$ is $\Delta = 0.2 \times \frac{dP(t)}{dV_{PV}(t)}$.

- **I-C method assisted by voltage-and-current-based ANN** [14] In this approach, the PV system only implements the I-C method when two neighboring power measurements does not exceed a given threshold, i.e., $|P(t) - P(t-1)| < h_1$. $h_1$ is tuned such that the false alarm period is also 100s. When the voltage-and-current-based ANN is triggered, the prediction of GMPP by ANN is given as a starting point of the I-C method.

- **I-C method assisted by irradiance-based ANN** [15] Here the ANN model utilizes the irradiance measurements from each PV panel. For a fair comparison, the irradiance-based ANN is triggered based the same threshold-based rule in [14]. The outputs of ANN are the minimum and maximum voltages ($V_{min}, V_{max}$) which form the boundaries of the global peak on the P-V curve (Fig. 4(b)). Whenever the ANN is triggered, the I-C method is forced to start with the predicted $V_{min}$.

*B. MPPT Performance in a Small PV System*

In the simulated small PV system, there are 12 PV panels each with size $5 \times 1$. The I-V and P-V characteristics under uniform irradiance (1000 W/m$^2$) are shown in Fig. 7(a) where the red dot indicates the maximum power $P_{MPP,0} = 5$KW with the corresponding voltage $V_{MPP,0} = 125$V. According to the input irradiance, the PV cells are grouped into three sets with size $5 \times 5$, $5 \times 5$ and $5 \times 2$ respectively. Two shading patterns ("SP 1" and "SP 2") are simulated by setting the received irradiance of three PV cell groups as {1000 W/m$^2$, 800 W/m$^2$, 500 W/m$^2$} or {1000 W/m$^2$, 300 W/m$^2$, 200 W/m$^2$}, and the corresponding I-V and P-V curves are given in Fig.7(b)-(c). The red dots denoting the location of the GMPPs reveals the optimal operating voltages leading to GMPP, i.e., $V_{GMPP,1} \approx 107$V and $V_{GMPP,2} \approx 50$V.

*1) Component-wise Improvements:* Tables. IV-VI demonstrate the advantages of three enhanced components where the results are all based on 500 independent experiments.

For the change detection component, as shown in Table IV, we evaluate the resource saving by the GLLR change detection that reduces the number of redundant triggers of the ANN. The saved resource is evaluated by the redundant ANN triggers within the time interval during which the system catches up with the new GMPP after the irradiance change. Specifically, we compared the average number of shading alarms per second $\overline{N}_{GLLR}$ with that of the threshold-based approach $\overline{N}_{th}$ given the same detection delay. The rate of resource saving is given by $\left(1 - \overline{N}_{GLLR}/\overline{N}_{th}\right) \times 100\%$. In particular, the rates of resource saving by our GLLR detector are

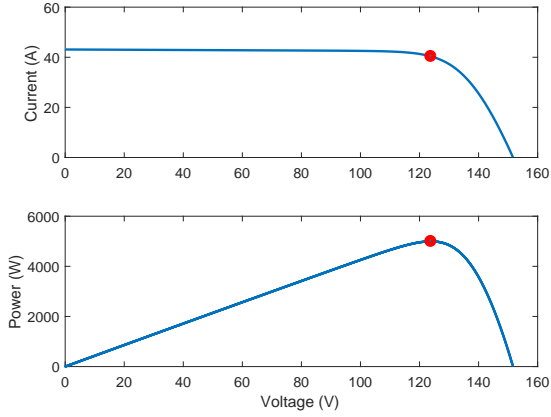

(a) I-V and P-V characteristics under uniform irradiance.

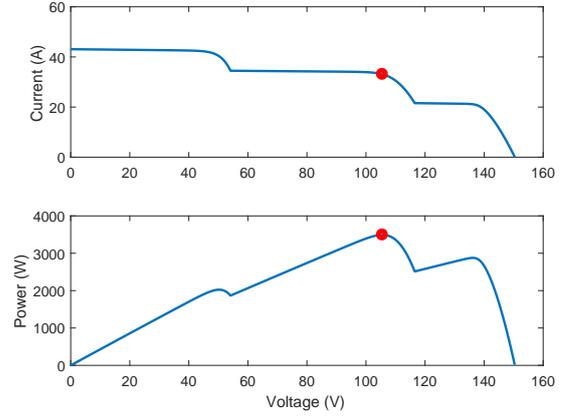

(b) I-V and P-V characteristics with SP 1.

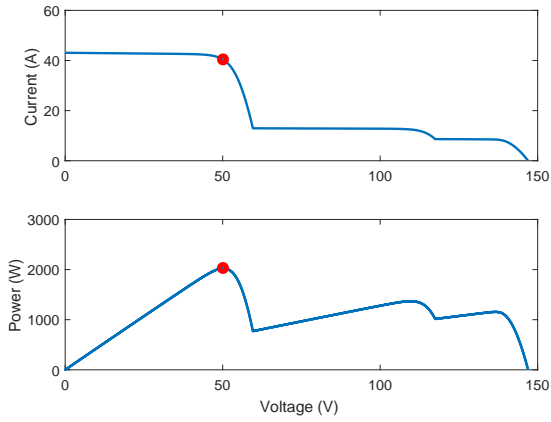

(c) I-V and P-V characteristics with SP 2.

Fig. 7. I-V and P-V characteristics of a small PV system under different irradiance conditions. The red dot denotes the global MPP point

all above 36.6% under different detection delays, which demonstrates the high efficiency of our sequential approach.

For the local MPP estimation component, we evaluate the average delay during which the estimated optimal operating voltage gradually approaches the new MPP, $P_{GMPP,2}$, after the shading pattern changes from "SP 1" to "SP 2". As shown in Table V, the delays of our proposed SMC are smaller than that of the I-C method in [15] [14].

For the ANN prediction component, Table VI shows the PQIs of our ANN models with multiple $(V_{PV}, I_{PV})$ or irradiance inputs, in comparison with the ANN model based on a single measurement. In general, the PQI values of our ANN models is 5% greater than that of the single-measurement based ANN models at least.

TABLE IV
IRRADIANCE CHANGE DETECTOR PERFORMANCE.

| Detection delay | Rate of resource saving | |
| --- | --- | --- |
| | "normal" → "SP 1" | "SP 1" → "SP 2" |
| 10 s | 36.6% | 42.6% |
| 15 s | 38.3% | 45.4% |
| 20 s | 38.1% | 46.3% |

TABLE V
MPP ESTIMATION PERFORMANCE (FROM "SP 1" TO "SP 2")

| Power | Delay | |
| --- | --- | --- |
| | SMC | I-C |
| $70\% \times P_{GMPP,2}$ | 0.06s | 0.08s |
| $80\% \times P_{GMPP,2}$ | 0.09s | 0.13s |
| $95\% \times P_{GMPP,2}$ | 0.22s | 0.35s |

TABLE VI
ACCURACY OF ANN PREDICTION (FROM "SP 1" TO "SP 2")

| Input type | PQI | |
| --- | --- | --- |
| | voltage and current based | Irradiance based |
| Multiple inputs | 96.2% | 94.5% |
| Single input | 91.1% | 87.7% |

*2) Overall Performance:* To present the efficiency of the proposed MPPT method, the input irradiances of three PV cell groups were tuned to switch the system mode ("normal" and various "shaded" modes). Fig. 8 shows the tracking dynamics of our method where the system receives uniform irradiance until $t = 5.75$s when the SP 1 occurs, and the SP 2 takes effect at $t = 7.75$s. In response to the abrupt irradiance change, the output power significantly decreases since the previous operating voltage no longer leads to the maximum power under different shading conditions. Then the system using our enhanced MPPT method quickly catches up with the new GMPPs within 0.7s while the delays of the methods in [14] [15] exceed 1s. In this small PV system, the performance difference between the voltage and current-based ANN and the irradiance-based ANN is not obvious.

We obtained the voltage tracking data in 500 identical experiments where the PV system mode switched from "SP 1" to "SP 2". Fig. 9 gives an average efficiency evaluation of the proposed

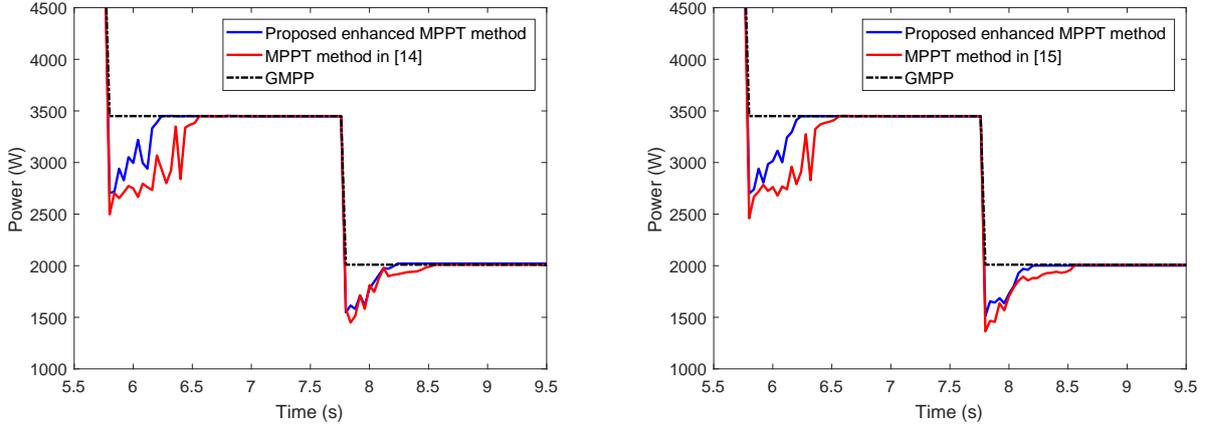

(a) Power dynamics with the ANN using voltage measurement.  (b) Power dynamics with the ANN using irradiance measurement.

Fig. 8. Power dynamics under different irradiance conditions in a small PV system. ($\gamma = 20$s, $\sigma_v = 10^{-5}$).

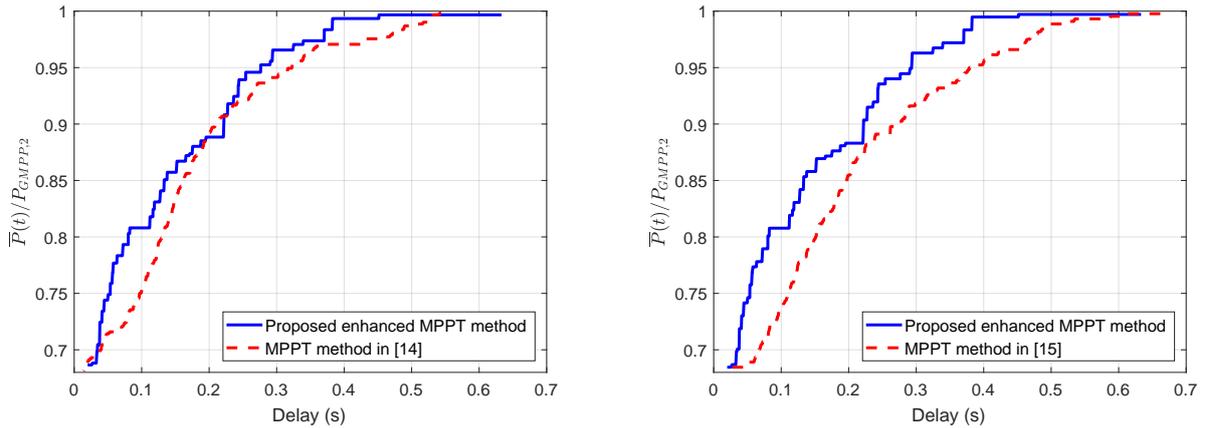

(a) Average efficiency performance with the voltage-and-current-based ANN.  (b) Average efficiency performance with the irradiance-based ANN.

Fig. 9. Average efficiency performance in a small PV system ($\gamma = 20$s, $\sigma_v = 10^{-5}$).

method. The horizontal axis denoted as "delay" refers to the time interval $(t - T_{shading})$ of 500 voltage measurements $\{P_i(t)\}_{i=1}^{500}$ taken at $t$ after the irradiance change at $T_{shading}$, and the vertical axis presents the corresponding value $\frac{\overline{P}(t)}{P_{GMPP,2}} \triangleq \frac{\sum_{i=1}^{500} P(t)}{500 V_{GMPP,2}}$. It is seen that the delay of the competing methods in [15] [14] is greater than 0.6s. In contrast, our proposed method can achieve $0.95 P_{GMPP,2}$ within 0.28s after the irradiance change and gain the new GMPP within 0.5s, and hence exhibits a higher efficiency.

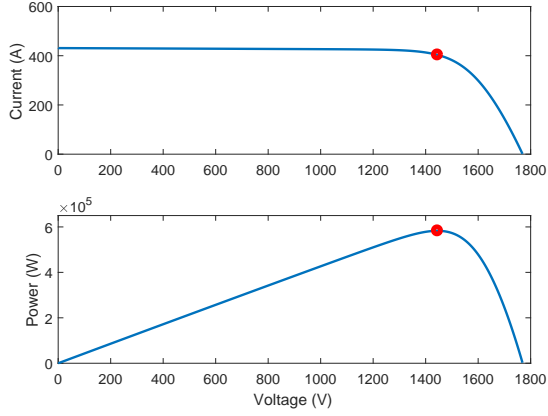

(a) I-V and P-V characteristics under uniform irradiance.

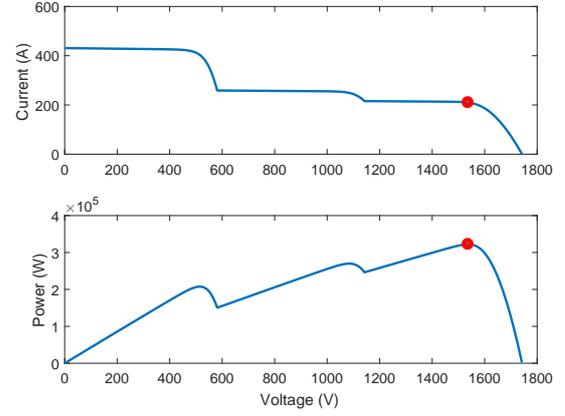

(b) I-V and P-V characteristics with partial shading pattern 1.

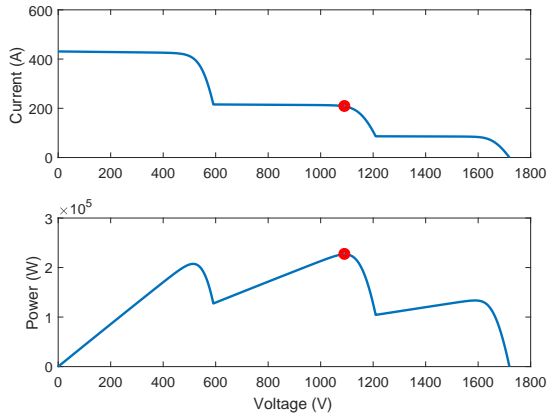

(c) I-V and P-V characteristics with partial shading pattern 2.

Fig. 10. I-V and P-V characteristics of a large PV system under different irradiance conditions. The red dot denotes the global MPP point

### C. MPPT Performance in a Large PV System

The large PV system consists of 120 PV panels each with size $50 \times 1$, and they are categorized into three groups of size $50 \times 50$, $50 \times 50$ and $50 \times 20$ respectively. As shown in Fig. 10(a), under uniform irradiance (1000 W/m$^2$), the maximum power is around 590KW and $V_{MPP,0} = 1400$V. Fig.10(b)-(c) presents the I-V and P-V characteristics under two shading patterns when the received irradiance of three PV cell groups are set as $\{1000$ W/m$^2$, 600 W/m$^2$, 500 W/m$^2\}$ or $\{1000$ W/m$^2$, 500 W/m$^2$, 200 W/m$^2\}$.

*1) Component-wise Improvements:* The performances of each component are given in Tables.VII-IX. Table VII presents the resource savings by the GLLR change detector with different detection delays. In comparison with Table IV, as the noise variance increases in the large PV system, the ANN is triggered more frequently and thus the amount of the saved resource slightly decreases. Our GLLR detector leads to less redundant alarms (smaller false alarms period) given the same detection delay. Table VIII shows that the SMC-based estimation achieves 95% of the new MPP within 0.28s, while the delay of the I-C method is 0.41s. The accuracy performance of ANN predictions is given in Table IX where our ANN model with multiple inputs has higher PQIs demonstrating the robustness of our method when the measurements are noisy. In general, the performance gaps between the enhanced components and their counterparts are more obvious than those in a less-noisy small PV system, which demonstrates the superiority of our proposed method.

TABLE VII
IRRADIANCE CHANGE DETECTOR PERFORMANCE.

| Detection delay | Resource saving | |
| --- | --- | --- |
| | "normal" → "SP 1" | "SP 1" → "SP2" |
| 10s | 34.6% | 41.6% |
| 15s | 37.3% | 42.4% |
| 20s | 37.6% | 43.3% |

TABLE VIII
MPP ESTIMATION PERFORMANCE (FROM "SP 1" TO "SP 2")

| Power | Delay | |
| --- | --- | --- |
| | SMC | I-C |
| $70\% \times P_{GMPP,2}$ | 0.08s | 0.11s |
| $80\% \times P_{GMPP,2}$ | 0.14s | 0.16s |
| $95\% \times P_{GMPP,2}$ | 0.28s | 0.41s |

TABLE IX
ACCURACY OF ANN PREDICTION (FROM "SP 1" TO "SP 2")

| Input type | PQI | |
| --- | --- | --- |
| | voltage and current based | Irradiance based |
| Multiple inputs | 94.2% | 92.3% |
| Single input | 85.4% | 83.2% |

*2) Overall Performance:* The power dynamics are shown in Fig. 11, where the irradiance condition first switches from "normal" to "SP 1" at $t = 5.8$s and then to "SP 2" at $t = 7.8$s.

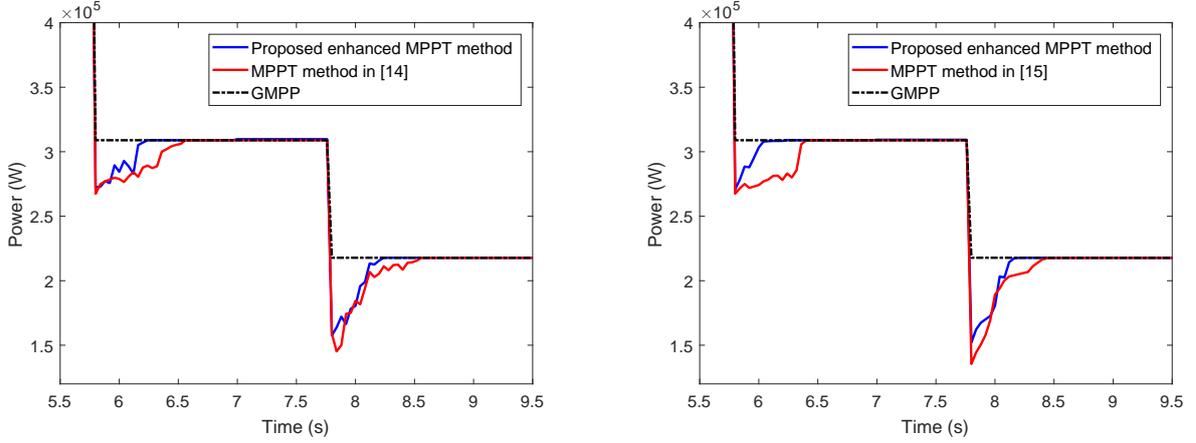

(a) Power dynamics with the ANN using voltage measurement. (b) Power dynamics with the ANN using irradiance measurement.

Fig. 11. Power dynamics under different irradiance conditions in a large PV system ($\gamma = 20$s, $\sigma_v = 10^{-3}$).

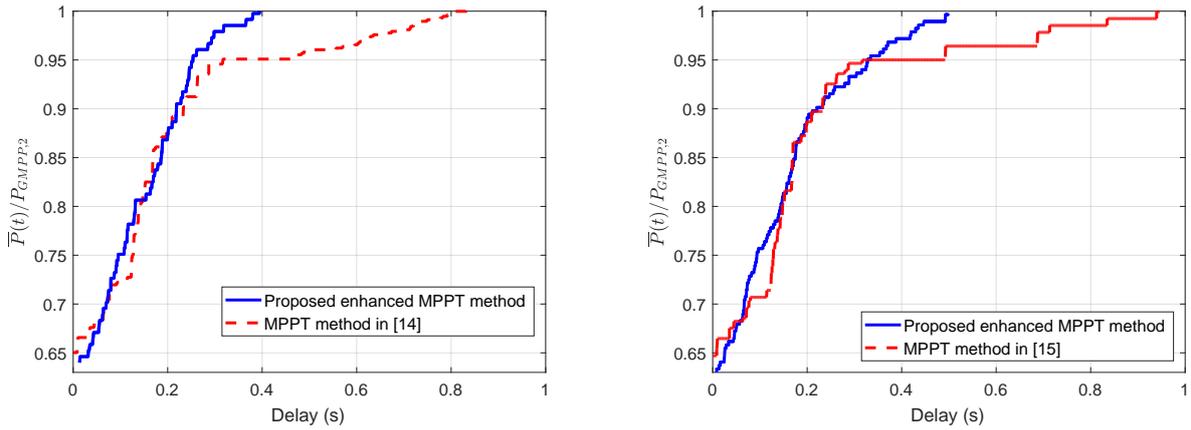

(a) Average efficiency performance with the voltage-and-current-based ANN. (b) Average efficiency performance with the irradiance-based ANN.

Fig. 12. Average efficiency performance in a large PV system ($\gamma = 20$s, $\sigma_v = 10^{-3}$).

The corresponding GMPP first decreases from 590KW to 340KV and then to 230KV. Similar to Fig. 8, the proposed MPPT method exhibits less delays ($\approx 0.4$s) before the the system arrives at the new GMPP.

Fig. 12 quantifies the average tracking efficiency. On one hand, compared with Fig. 9, the delay of our methods almost maintains the same as that in the small system, while the delays of the methods in [15] [14] increase from 0.6s to 0.85s. The obvious gap between the two implies the robustness of our method.

## V. Conclusion

In this paper, we proposed an improved MPPT method for PV systems under partial shading, by utilizing the prediction of the GMPP via ANN to refine the SMC-based I-C tracking algorithm. The SMC algorithm tackles the nonlinear voltage transition model when the step size is variable. The ANN model is based on the input of the voltage and current or the irradiance measurements, and predicts the GMPP given the knowledge learned from training data. Furthermore, the ANN-based refinement is triggered only when the proposed GLLR change detector declares the irradiance change, which decreases the number of redundant ANN predictions when the irradiance is steady. The simulation results demonstrates that our proposed method can efficiently track the MPP and is robust to various partial shading patterns and different measurement noises.